\documentclass[a4paper,12pt]{article}
\usepackage{graphicx, subcaption, setspace, siunitx}
\usepackage{amsmath,amsfonts,amsthm, amssymb, textcomp}
\usepackage{dsfont}
\usepackage{setspace}
\usepackage[numbers]{natbib}
\usepackage{bibref}
\usepackage[utf8]{inputenc}
\usepackage[english]{babel}
\usepackage{xcolor}
\usepackage{gensymb}
\usepackage{hyperref}
\usepackage{multirow}
\usepackage{url}
\usepackage{physics, tensor}
\usepackage{listings, epigraph}

\hypersetup{colorlinks,linkcolor=red!50!black,citecolor=blue, urlcolor=cyan, anchorcolor=sepia}

\author{Julian De Vuyst}
\date{\emph{Department of Physics \& Astronomy,\\
Ghent University, Krijgslaan, S9, 9000 Ghent, Belgium}}
\title{ {\Huge \bf A natural introduction to Fine-Tuning}}

\graphicspath{{Pictures/}}

\AtBeginDocument{}

\begin{document}
\maketitle

\begin{abstract}
    A well-known topic within the philosophy of physics is the problem of fine-tuning: the fact that the universal constants seem to take non-arbitrary values 
    in order for life to thrive in our Universe. In this paper we will talk about this problem in general, giving some examples from physics. We will review some solutions like the design argument, 
    logical probability, cosmological natural selection, etc.~Moreover, we will also discuss why it is dangerous to uphold the Principle of Naturalness as a scientific principle. 
    After going through this paper, the reader should have a general idea what this problem exactly entails whenever it is mentioned in other sources.
\end{abstract}

\newpage
\tableofcontents
\newpage

\section{Introduction}
\indent Physics still remains the best framework to describe Nature where possible (the motion of planets, the distribution of heat, etc.). It is built from intricate models each describing a specific part of reality in a specific regime. These models can be reached from one another in certain limits, e.g. 
classical physics can be reached from special relativity if one takes the limit of the speed of light going to infinity. Such models often require a set of free parameters, and different choices of values for these parameters lead to physically different behaviour.
Therefore, for a model to describe our reality as closely as possible one needs to perform measurements. Often it turns out that the value is an extremely large or extremely small number without a good reason for it to be so large/small.
In this sense, physicists like to invoke the principle of naturalness to prefer models which have parameters of order 1, but such models might not describe our reality. This leads to the fine-tuning problem: why do the constants take such values in order for 
our Universe to properly function as it does. Whilst many philosophers and physicists have proposed answers, from easy to intricate ones, this question remains unsolved. Mostly due to the fact that any solution turns out to be unobservable and even if it was, we only have our Universe, hence making 
such measurements have a sample size of only one.\\
\\
\indent This short paper is meant to introduce the interested reader to this problem and along the way hopefully clarify why this problem is hard to solve. 
The paper is structured as follows: in section \ref{section2} we discuss the importance of constants and their dimension, this is followed by section \ref{section3} in which we will introduce the concept of naturalness. 
Subsequently, section \ref{section4} describes the problem itself and in section \ref{section5} we look at some examples of finely tuned parameters. In section \ref{section6} we look at some proposed solutions. 
Afterwards, section \ref{section7} discusses the fading of the barrier between science and art when using naturalness as a guiding principle. At last, some conclusions are presented in section \ref{section8}.\\
\indent The reader is not required to have an extensive knowledge of physics. We will try to explain new concepts in a coherent, easy to understand way so that this paper may stand on itself and does not require the aid of some textbook.

\section{A take on constants}
\label{section2}
\indent Parameters, constants, and variables arise in all physical models of nature e.g.~$\pi, \hbar, c$... They are ubiquitous in equations and show up due to certain other theoretical or experimental relations 
used to derive those expressions. While variables are something which we control, we can e.g.~measure the time it takes for a ball to fall over some variable distance, constants are 
fixed quantities out of our control. They take a certain value and that is it, nothing can be done about it unless they are dimensionful and/or quantum effects play a role.

\subsection{The role of units}
\indent Constants and variables are both usually expressed in some type of units, for instance time can be expressed in seconds, hours, years, etc.~\cite{Wall2016} 
These units are called the \emph{dimension} of the quantity and dimensional analysis plays a crucial role in physics. However, dimensionful constants on their own are meaningless,\footnote{From now on we will restrict our focus to constants, as these will play the prominent role in this paper. Hence whenever we mention parameter, it should be thought of as a constant. Note that the following can also be said for 
variables, but in this case units are necessary to relate the results to our physical world so that they do have meaning.}
because the numerical value they take depend on the units we chose to express the quantities in. As an example we can look at the reduced Planck's constant expressed in two different units:
\begin{align}
    \hbar = 1.055 \cdot 10^{-34}~\text{J s} = 6.582 \cdot 10^{-16}~\text{eV s}.
\end{align}
The large difference of the order of $10^{18}$ is purely due to a rescaling of our chosen unit of energy: $1~\text{J} = 6.242 \cdot 10^{18}~\text{eV}$.\\
\indent Therefore, for any dimensionful constant there exists a set of units such that its numerical value can be rescaled to a value of order $1$.
This is often used in physics to ease certain large computations or to save processing power during numerical simulations. For example, in high energy physics it is customary to set the reduced Planck's constant $\hbar$, the speed of light $c$, and Boltzmann's constant $k$ 
equal to one.\footnote{This particular system is called \emph{natural units}.}~Consequently, this sets the values for the time scale, length scale and temperature scale respectively in terms of the energy. In the end, when using dimensional analysis one can always find exactly one combination of the powers of these constants $ \hbar^\alpha c^\beta k^\gamma$ 
such that the equations are in our desired units. This rescaling is what makes their numerical values meaningless.\\
\\
\indent On the other hand, there are also constants which have no units and hence are called \emph{dimensionless}: $\pi$ being an obvious example. These constants are just numbers independent of 
a certain set of units we chose to describe our physical system. Therefore, these numbers do have meaning because we cannot rescale them to any other value we want. 
This concept of dimensionlessness will play an important role in naturalness and the fine-tuning problem.\\
\indent Note that it is always possible to construct a dimensionless parameter from two dimensionful parameters with the same dimension. One simply has to take the ratio 
of these two dimensionful constants such that the dimension drops out. For example, the ratio of the proton mass to that of the electron $m_p/m_e$ is dimensionless.

\subsection{Derived vs fundamental constants}
\indent There is another subdivision within constants which is related to how their value is determined \cite{Wall2016}. Constants whose value can be deduced from other deeper facts about physics and mathematics are called \emph{derived} constants. Such constants include $\pi$ which 
is derived from the ratio of the circumference of a circle and its radius, and $g$ which can be derived from Newton's gravitational law close to the Earth's surface. Note that, while there are literally 
dozens of ways to calculate $\pi$, its numerical value does not depend on other constants. This is in contrast with $g$ whose value depends on the radius of the Earth, its mass and Newton's constant.\\
\\
\indent In the other case, a \emph{fundamental} constant, often called a free parameter, is a quantity whose numerical value cannot be determined by any computations. In this regard, it is the lowest building block of equations as these 
quantities have to be determined experimentally. The Standard Model alone contains 26 such free parameters \cite{Thomson2013}:
\begin{itemize}
    \item The twelve fermion masses/Yukawa couplings to the Higgs field;
    \item The three coupling constants of the electromagnetic, weak and strong interactions;
    \item The vacuum expectation value and mass of the Higgs boson;\footnote{See \ref{Higgs}.}
    \item The eight mixing angles of the PMNS and CKM matrices;
    \item The strong CP phase related to the strong CP problem.\footnote{See \ref{CP}.}
\end{itemize}
Other free parameters include Newton's constant, the cosmological constant\footnote{See \ref{cosm}.}, and so forth.\\
\indent As these free parameters do not come from any deeper theory of physics, yet, their numerical values are such that they match the observations as closely as possible.
In this light, free parameters and their derived constants contain error bounds due to lack of precision or due to different results from different methods of measurement. 
On the contrary, mathematically derived constants can be exactly calculated and do not have an error associated with them.\\
\\
\indent In conclusion, when talking about naturalness and fine-tuning we are talking about \textbf{dimensionless}\footnote{We will see that this is not always strictly true, but it can always be recast in such a form.} and \textbf{fundamental} constants, because 
\begin{enumerate}
    \item These are the constants with a fixed value regardless of the model we are considering and therefore have significance;
    \item These constants cannot be derived from other constants and have to be verified by experiment. Simply put: we do not know why they have that value.
\end{enumerate}

\section{The concept of naturalness}
\label{section3}
The concept of naturalness has changed its meaning throughout its evolution \cite{borrelli2019practice, hossenfelder2019screams}. When physicists today talk about naturalness, it will mostly refer to numerical naturalness.

\subsection{Technical naturalness}
\subsubsection{A Wilsonian perspective}
\begin{figure}[h!]
    \centering
    \includegraphics[width=\textwidth, clip]{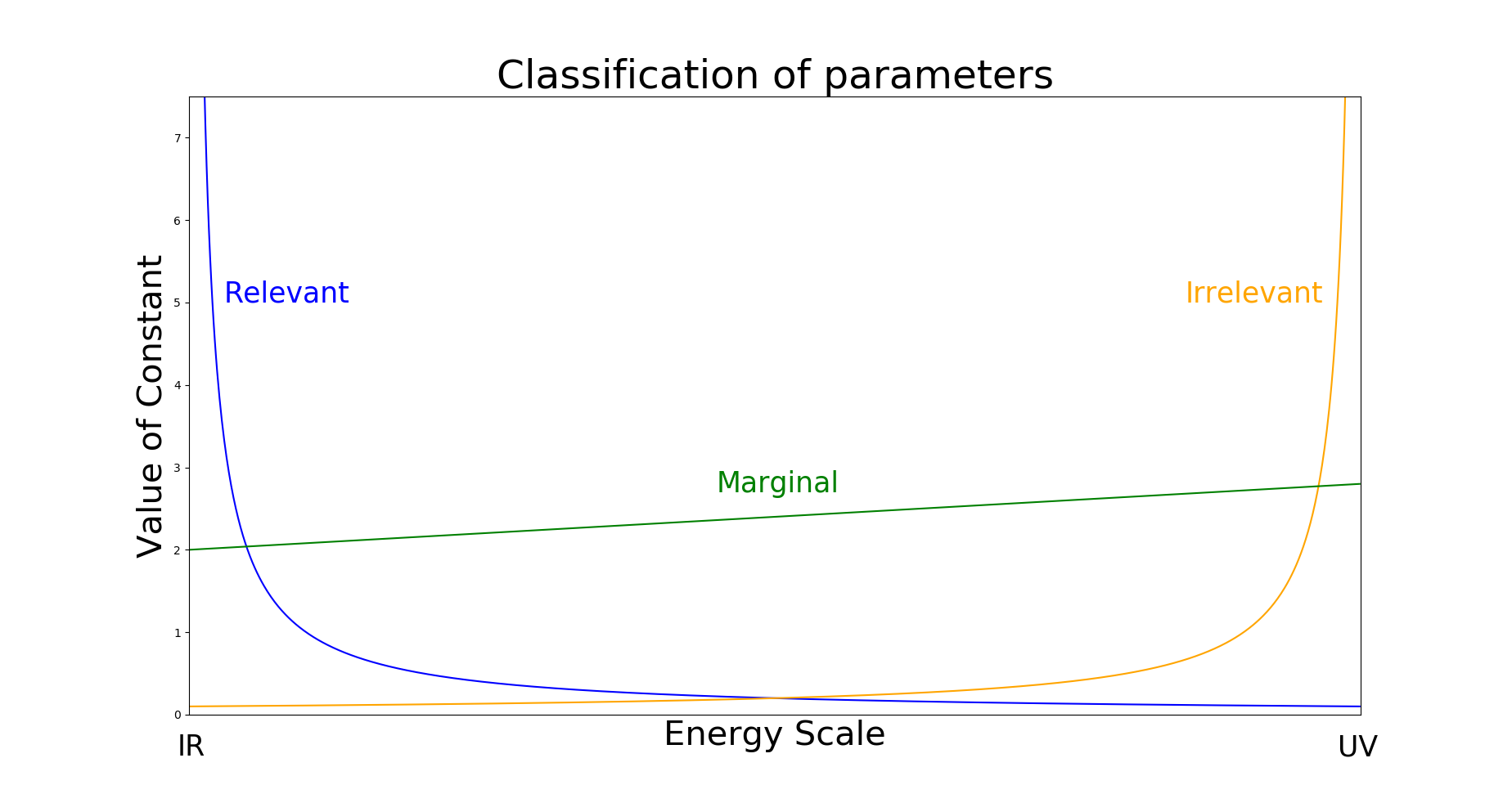}
    \caption{Classification of the three parameters according to the Wilsonian perspective.}
    \label{Wilson}
\end{figure}

\indent The usual beginning of naturalness has been marked by K.~Wilson's famous paper \cite{Wilson1971} introducing the Renormalisation Group (RG) method and its physical meaning \cite{borrelli2019practice}. 
In short, this paper makes apparent that the laws of nature are size dependent which means that the constants are dependent on the size/scale at which we measure them. How a constant exactly behaves under a change 
of scale is what is called the \emph{RG flow} and constants are classified according to their flow from high energy/UV/short distance scales to low energy/IR/long distance scales where they constitute what physicists call a 
\emph{low energy effective field theory}. In the Wilsonian approach, one introduces a cut-off scale to integrate out the degrees of freedom laying beyond this cut-off. This is analogous to getting a macroscopic theory from a microscopic one, an example of this 
is how hydrodynamics can be understood as a microscopic theory of interacting particle.\\
\indent Two things should be noted: 
\begin{itemize}
    \item The phenomenological parameters describing the macroscopic theory are clearly derived as they are calculated from the microscopic theory; 
    \item A changing value due to a changing scale implies these parameters have a certain dimension.\footnote{Again, we will see that this is not strictly true.}
\end{itemize} 
According to their behaviour under the RG flow, we can distinguish three types dependent on their energy dimension $E^\alpha$ with $\alpha \in \mathds{R}$:\footnote{To determine this dimension, one generally assumes natural units.}
\begin{enumerate}
    \item \emph{Relevant} parameters are those that acquire larger values when lowering the energy scale. This implies they have units of $E^{\alpha<0}$;
    \item \emph{Marginal} parameters do not or barely change under a flow to the IR, this requires $E^0$, making them dimensionless;
    \item \emph{Irrelevant} parameters tend to vanish in the macroscopic theory, hence they are required to have units of $E^{\alpha>0}$.
\end{enumerate}
This behaviour can be summarised in Figure \ref{Wilson}.

\indent This view on physics implies a certain sense of scientific reductionism \cite{Wall2016} as the behaviour of physical systems seems to be determined by the laws of nature on the smallest parts and we do not regard 
any consequences of complexity. However, this view is actually wrong due to the concept of \emph{emergence}: this gives a holistic view on physics, namely that macroscopic theories cannot be fully described by the interactions at 
microscopic scales since due to the high number of degrees of freedom involved some new phenomena may occur in the macroscopic theory not present in the microscopic one. A de facto example is the weather: although the equations and principles are not that hard to understand, together 
they tend to a highly non-trivial system which is hard to simulate.\\
\\
\indent According to our previous section, naturalness would only apply to marginal parameters since they are effectively dimensionless. This should raise two questions: how can a dimensionless parameter's value 
change under RG flow and what about (ir)relevant parameters. The answer is simply put: quantum effects \cite{Wall2016, borrelli2019practice, hossenfelder2019screams}. Going to lower energy scales also determines which particles and forces tend to activate or deactivate
at that level. The contribution of these new `sources' can effectively change the value of the parameters,\footnote{Although some parameters can be protected due to a symmetry principle. We will not go deeper into this.} whether they are relevant, irrelevant or marginal.\\
\indent In this respect one defines a parameter to be \emph{technically natural} \cite{Wall2016, borrelli2019practice, hossenfelder2019screams} if its value in the IR is nearly independent of its value in the UV. Since in most cases we do not know the theory in the UV, the parameters could have any value and the choice does not affect what happens at low energies. 
It reflects the idea that we should not cherry-pick our parameters and that a theory at low energies is likely even if we had randomly picked the parameters at high energies.\footnote{Note that this is a purely mathematical change, not one which may actually happen.}
This is not the same as saying the IR and UV theories decouple, as the first is still dependent on the latter. We just do not have to know about what happens at the smallest distances to accurately predict phenomena at large distances.
Conversely, technically unnatural numbers would therefore be constants for which their value in the UV has to be finely tuned in order to have a certain value in the IR.\\
\indent Examples of technically unnatural values include the mass and self-energy of the electron, the difference of pion masses, the absence of flavour changing neutral currents, etc. 
This last problem was resolved by introducing the charm quark and is the only prediction ever made on technical naturalness. Three failures which we will come back to include the cosmological constant, the Higgs boson mass, and the strong CP problem.

\subsubsection{A brief history}
According to \cite{borrelli2019practice} the history can be structured into four phases:
\begin{enumerate}
    \item \textbf{'70-'79}: during the emergence of the Standard Model, the methods of the RG group and spontaneous symmetry breaking were explored. These methods were used to try and predict the value of masses and coupling constants.
    During this phase, `natural' had another definition: it occurred when a relation between two parameters only received finite radiative corrections due to spontaneous symmetry breaking;
    \item \textbf{'80-'85}: the Standard Model was fully established and physicists were looking for a more unified theory. This was mainly guided by vague problems which were called `naturalness problems'. For instance, the Higgs mass was such a problem. Naturalness was hence used as a conceptual tool for producing problems to solve. During this phase 
    the term `natural' became popular. Different definitions were used but they all gave a solution to some problem with the Standard Model, it was mainly used to promote a specific model over another;
    \item \textbf{'85-2012}: LEP results failed to prove supersymmetry\footnote{See \ref{Higgs}.} which was the most promising solution to the naturalness problem at that time. There came critique from P.~Nelson \cite{Nelson1985} that theoretical physics was less determined by experiment than before: naturalness became a principle for 
    formulating theories. Not only was it a guide to choose among theories, but it was also used to point at problems and prompted the search for new theories to solve them. His article made claims about what has now become numerical naturalness. Technical naturalness also came into use. Hence, we see that the goals of research shaped the criterion of naturalness;
    \item \textbf{2012- }: the results of the LHC confirm the predictions made by the Standard Model but show no evidence of new physics. It reheated the discussion about naturalness.
    
\end{enumerate}

\subsection{Numerical/General naturalness}
\indent When talking about naturalness today, most physicists refer to the notion of \emph{numerical naturalness} \cite{hossenfelder2019screams}: a theory's dimensionless parameters should not be much larger or smaller than 1.
Note that small numbers and large numbers are not treated as two separate cases as one can easily convert the first to the latter or vice versa by taking its inverse. Thus, taking the difference of two 
constants may also produce a number deemed unnatural as the difference may be very small compared to the original absolute values.\\
\indent Note that our previous definition of a technically unnatural number coincides with a numerically unnatural number in the UV. So, essentially it all boils down to numerical naturalness.\\
\\
Before we continue, we present a short recap of both views:
\begin{enumerate}
    \item \textbf{Technical Naturalness} generally deals with both relevant, \textbf{dimensionful} and marginal, \textbf{dimensionless} parameters;
    \item \textbf{Numerical Naturalness} generally deals with \textbf{dimensionless} parameters;
    \item Both views consider \textbf{fundamental} parameters;
    \item Technical unnaturalness can imply numerical unnaturalness;
    \item Dimensionful parameters can always be combined to create dimensionless parameters.
\end{enumerate}

\section{The Fine-Tuning Problem}
\label{section4}
\begin{figure}[h!]
    \centering
    \includegraphics[width=\textwidth, clip]{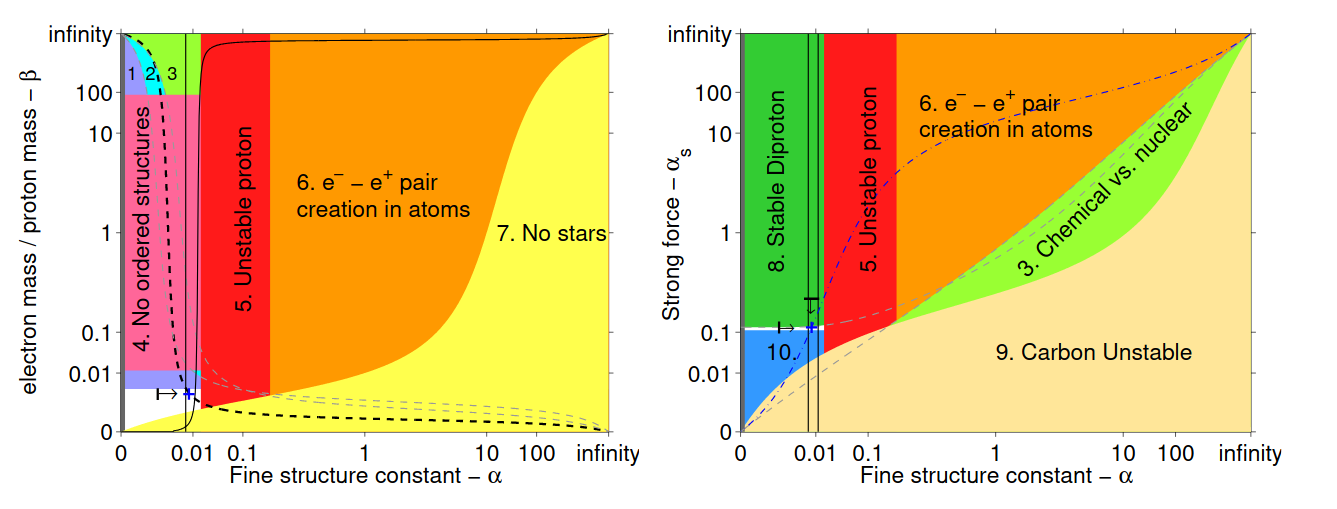}
    \caption{the demarcation of the ratio $m_e/m_p$ and the constant $\alpha$ to permit life in our Universe. The life-permitting region is shown in white and our Universe 
    is represented by the blue cross. Taken from ``The Fine-Tuning of the Universe for Intelligent Life''\cite{Barnes2012}.}
    \label{Goldilocks}
\end{figure}

\indent Fine-tuning is related to the numerical values of certain fundamental constants and can be applied to both technical and numerical unnatural numbers. We have already seen that a certain constant may need to be finely tuned in the UV to 
become a certain value in the IR. Closely related to this problem is what P.~Davies describes as the \emph{Goldilocks Principle}: for life to emerge in our Universe, the fundamental constants could not have been more than a few percent from their actual values (see Figure \ref{Goldilocks})\cite{Davies2007}.
For instance, for elements as complex as carbon to be stable, the electron-proton mass ratio $m_e/m_p = 5.45 \cdot 10^{-4}$ and the fine-structure $\alpha = 7.30 \cdot 10^{-3}$ may not have values differing greatly from these \cite{Colyvan2005}. They are just about right.
The study of such relations is referred to as \emph{anthropic fine-tuning}: the quantification of how much these parameters in our theories could be changed so that life as we know it would still be possible \cite{hossenfelder2019screams}.
As we will see, this leads to the design argument and the anthropic principle as possible solutions. Notice that this kind of fine-tuning is different from the one described previously, whether a constant takes on a natural value is not of importance in this case. \\
\indent Even so, the Goldilocks principle is not a scientifically proven fact as it heavily depends on what we define as `life'.\footnote{My thanks to Johan Braeckman for pointing this out.} Life as we know it is carbon-based, we have yet to encounter another species with different properties of fulfilling what life is according to their inner workings. Silicon-based life 
could be possible for instance \cite{doi:10.1126/science.aah6219}. 
In this regard, our definition of life seems to be Popper falsifiable. But, as our sample size of life-thriving planets is currently just one we cannot know for sure if there are other forms of life possible. Nevertheless, G.~Lewis and L.~Barnes argued that according to fine-tuning considerations, 
universes with different constants would give rise to less structure and complexity. This would rather make such universes life-hostile, regardless of how we define `life' \cite{Friederich2017, Lewis}.\\
\indent Still, the problem remains the same:
\begin{center}
    \textbf{The Crux}: why are the fundamental constants as they are.
\end{center}

\section{Some examples from physics}
\label{section5}
In this section we will briefly describe four common finely tuned constants.

\subsection{The Cosmological Constant Problem}
\label{cosm}
During the development of the theory of General Relativity, A.~Einstein derived the following equation describing the relation between the geometry of spacetime ($g^{\mu \nu}, R^{\mu \nu}, R$) and the energy/matter distribution  ($T^{\mu \nu}$) in that spacetime \cite{Rijcke}:
\begin{equation}
    R^{\mu \nu} - \frac{1}{2} R g^{\mu \nu} = -\frac{8\pi G_\text{N}}{c^4} T^{\mu \nu},
\end{equation}
where $G_\text{N}$ is Newton's constant and $c$ is the speed of light.\\
\indent Observations concluded that the Universe at cosmological scales, at the scale of galaxy clusters, is homogeneous and isotropic. Furthermore, the Universe seems to be in an era of expansion (first noted by E.~Hubble and G.~Lema\^itre). 
However, this expansion is not embedded in the Einstein equation described above. In order to obtain an expanding universe, one can add an extra term to the left-hand side of the Einstein equation: $\Lambda g^{\mu \nu}$ where $\Lambda$ is deemed the 
\emph{cosmological constant}.\footnote{Its original introduction was to get rid of the boundary conditions for open universes \cite{Rijcke}.} In this interpretation, $\Lambda$ acts as a constant vacuum energy density of some fluid permeating our Universe notoriously dubbed \emph{dark energy}.\\
\indent It turns out that when we apply quantum field theory to compute this vacuum energy contribution, it is proportional to $M_\text{top}^4$ where $M_\text{top} = 10^{11}~\text{eV}$ is the mass of the top quark, the currently heaviest particle in the Standard Model \cite{hossenfelder2019screams}. 
At the end of the day, the theoretically obtained value and the experimentally observed one differ by 121 orders of magnitude.\footnote{This is often regarded as `the worst theoretical prediction in the history of physics'.} Even when one takes supersymmetry to be true,\footnote{See section \ref{Higgs}.} the 
difference still remains at about 80 orders of magnitude.\\
\indent What we have not yet taken into account is the quantum nature of this vacuum density, namely that there are quantum corrections $\lambda'$ to the bare value $\lambda$, the value proportional to $M_\text{top}$, rendering the cosmological constant so small. 
This is the same procedure as in the renormalisation of a quantum field theory where this extra contribution is itself not observable such that these `infinities' can be subtracted from each other leaving a finite value to be measured \cite{hossenfelder2019screams}.
We can hence write 
\begin{equation}
    \Lambda = \lambda + \lambda'.
\end{equation}
To get a dimensionless parameter, we can divide this equation by $\lambda$:
\begin{equation}
    \frac{\Lambda}{\lambda}  = 1 + \frac{\lambda'}{\lambda}.
\end{equation}
Since $\Lambda \ll \lambda$ this means that $1+\lambda' / \lambda$ is extremely small and this is \emph{the Cosmological Constant Problem}.\\
\indent Some theories which try to resolve this issue are the consideration of an inhomogeneous metric \cite{Wang2020}, time-dependent scalar field expectation values, sub-universes due to fluctuations in this scalar field, 
first-order phase transitions, quantum measurements, modified gravity, etc. \cite{weinberg1989cosmological}.

\subsection{The Flatness Problem}
\begin{figure}[h!]
    \centering
    \includegraphics[width=.5\textwidth, clip]{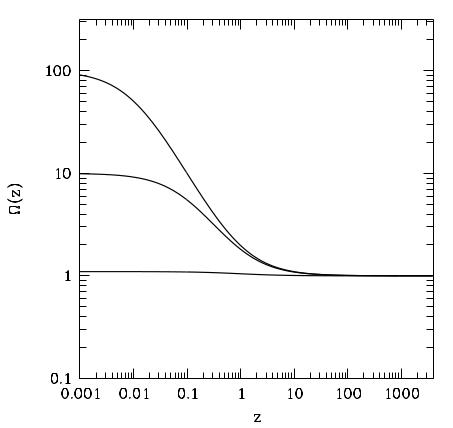}
    \caption{The evolution of different initial $\Omega_\text{early}$ in function of the redshift $z$. This redshift quantifies the lookback time: a higher $z$ corresponds to 
    earlier times in our Universe. Taken from \cite{Rijcke}.}
    \label{flatness}
\end{figure}

It is well-known that our expanding Universe can be well-described by a Friedmann-Lema\^{i}tre-Robertson-Walker (FLRW) metric which is the general metric for an isotropic, homogeneous expanding universe.
When looking at the first Friedmann equation one can bring this in the following form \cite{Rijcke}
\begin{equation}
    \frac{kc^2}{R(t)^2} = H(t)^2 [ \Omega(t) - 1],
\end{equation}
in which $k$ is the curvature of our spacetime, $R$ is the scale factor describing how the spatial part of our metric depends on time, $H$ is the Hubble parameter 
describing how the scale factor evolves, and $\Omega$ is the total density (matter and radiation) divided by some critical density, hence it is dimensionless. \\
\indent Our Universe today is very flat, meaning that 
$k \approx 0$. Hence, from the equation follows that $\Omega$ today is very close to $1$. Using some algebraic manipulations one can compute what $\Omega$ was in the early Universe. It turns out that to have 
$\Omega = 1$ today, $\Omega_\text{early}$ must also lie extremely close to 1. Small variations in $\Omega_\text{early}$ lead to big deviations from $\Omega = 1$ as seen in Figure \ref{flatness}.
This fine-tuning of $\Omega_\text{early}$ is called \emph{the Flatness Problem}. This problem is what \emph{inflation theory} solves.

\subsection{The Higgs mass}
\label{Higgs}
\begin{figure}[h!]
    \centering
    \includegraphics[width=.7\textwidth, clip]{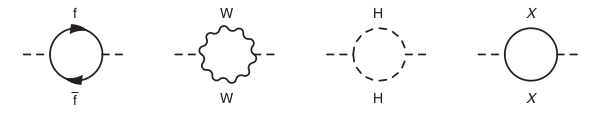}
    \caption{Some loop diagrams, Feynman diagrams, contributing to the mass of the Higgs boson $H$ (represented by a dashed line). $X$ represents a new massive particle. Taken from \cite{Thomson2013}.}
    \label{loops}
\end{figure}

The Higgs boson, only just recently discovered, plays an important role in the Standard Model: it is the only scalar/spin-0 field and the coupling to this boson is responsible for a part of the mass of the elementary particles.
A problem with this particle is that its (bare) mass $m_0^2$ receives quantum corrections from loop diagrams \cite{hossenfelder2019screams, Thomson2013}, see Figure \ref{loops}.\footnote{This is the same mechanism as in Section \ref{cosm}.} If our current understanding of the Standard Model indeed breaks down at energy scales 
of the order $\Lambda = 10^{16}-10^{19}~\text{GeV}$ this implies that these corrections are very large because they are themselves proportional to $\Lambda^2$:
\begin{equation}
    m_H^2 = m_0^2 + \sum_i c_i \Lambda^2 = m_0^2 + m_\text{loops}^2,
\end{equation}
where $c_i$ are some dimensionless coefficients.\\
\indent To recast this into a dimensionless parameter, one can simply divide the whole equation by $m_0^2$:
\begin{equation}
    \frac{m_H^2}{m_0^2} = 1 + \frac{m_\text{loops}^2}{m_0^2},
\end{equation}
the problem then resides in $m_\text{loops}^2/m_0^2$.\\
\indent To keep the Higgs mass within the observed range one hence requires large cancellations among each other in the loop corrections, effectively introducing a new term which almost exactly cancels the loop term to a high degree of precision. This is no problem as 
this new term is not observable and the current value of the Higgs mass agrees well with predictions and measurements.
This is dubbed \emph{the Hierarchy Problem} since these extra terms need to be fine-tuned.\\
\indent A popular solution to this problem is supersymmetry: the statement that each particle in the Standard Model has a superpartner called an sparticle. These new sparticles can introduce new loops effectively countering the original large contributions.
Yet, as these loops depend on the masses of these sparticles, any sparticles with a mass higher than the Higgs mass would reintroduce the problem. The problem is hence shifted from arbitrarily precise loop cancellations to the masses of these sparticles.
Nevertheless, there is currently no experimental evidence for supersymmetry.

\subsection{The Strong CP Problem}
\label{CP}
For any quantum field theory to be consistent, it must be symmetric under the combined action of CPT \cite{Thomson2013}:
\begin{enumerate}
    \item C-symmetry (charge conjugation) states that the fundamental laws for particles are the same for their antiparticles. Basically obtained by switching the sign of the particle charges;
    \item P-symmetry (parity) states that the system is invariant under a point reflection about the origin $\vec{r} \to -\vec{r}$. The laws of Nature are hence the same in a mirrored Universe;
    \item T-symmetry (time reversal) states that a system is invariant under time reversal. For instance, the Newtonian equations are valid for time running forwards as well as for time running backwards.
\end{enumerate}
It is well-known that the weak interaction violates CP-symmetry, hence differentiating on particle/antiparticle and what physicists call left/right chirality. It also violates T-symmetry such that the combined transformation CPT remains a valid symmetry.\\
\indent However, the strong interaction can in theory violate CP-symmetry, but it turns out it does not. No experiment has ever been performed in which this violation is seen. CP-violation is modelled into the Lagrangian as the \emph{$\theta$-term} 
and it turns out that for the strong interaction this parameter is vanishingly small: $\theta_\text{CP} < 10^{-10}$. Among physicists, this is known as \emph{the Strong CP Problem}.\\
\indent A proposed solution is to promote $\theta_\text{CP}$ to a dynamical field describing a particle called the \emph{axion} \cite{hossenfelder2019screams, Thomson2013, Weinberg1978, Wilczek1978}.
Although this particle is experimentally ruled out \cite{Dicus1978, Dicus1980}, a new model has been proposed with another particle called the \emph{invisible axion} and some physicists believe this could be the elusive dark matter.

\section{Resolutions to fine-tuning}
\label{section6}
Let us give an overview of a few proposed solutions to the fine-tuning problem, starting with the anthropic principle.

\subsection{We are here, full stop}

\emph{The Anthropic Principle} reminds us to take the effect of observation selection into account and often comes in three types \cite{weinberg1989cosmological, Carter74}:
\begin{enumerate}
    \item \textbf{Very Weak}: we are here as one experimental datum itself. For example, we `know in our bones' that the lifetime of the proton is larger than $10^{16}$~years, otherwise we would not survive the ionising particles produced by its decay;
    \item \textbf{Weak}: the constants are what they are, because it permits life;
    \item \textbf{Strong}: the laws of Nature are such that intelligent live can arise capable of observing those laws; they are completed by this requirement. Science is meaningless without observers. 
\end{enumerate}
Note that the weak and strong version differ in an important property: the strong version states that the laws are the result of the emergence of life, this is a necessary requirement, whilst the weak version simply states it allows for life. The very weak version is not that interesting, because it does not explain anything 
and does not give any relevant information. The strong anthropic principle seems unreasonable: science is obviously 
impossible without scientists, but the Universe is not necessarily impossible without scientists \cite{weinberg1989cosmological}.\\
\indent The weak principle is by far the most interesting one, because it allows us to construct quantitative arguments.
It could explain what eras and parts of the Universe we inhabit by calculating which eras and parts we could inhabit. An interesting example by dimensional analysis is the combination
\begin{equation}
    \frac{\hbar}{G_\text{N} c m_\pi^3} = 4.5 \cdot 10^{10}~\text{years},
\end{equation}
where $m_\pi$ is the pion mass.\\
\indent Note how this is roughly of the same order of the age of our Universe \cite{weinberg1989cosmological}! R.~Dicke pointed out that the question of the age of the Universe can only arise when the conditions are right for the existence of life. Hence, the Universe must be old enough 
so that some stars have completed their time on the main sequence and result in the production of heavy elements, and it must be young enough so that stars would still provide energy through nuclear reactions. Both the lower and upper bounds can be very roughly given by that quantity.
This principle can be used in the cosmological constant problem since the Universe does contain eras and parts during which the effective cosmological constant takes a wide range of values. \\
\indent At best, the weak version cannot explain why certain constants take exactly their value, but it can provide us with upper and lower bounds based on reasonable assumptions what any other theory could do as well.


\subsection{Designed like clockwork}

\indent Just like the intricate workings of a watch, could the Universe and its finely tuned constants also be created by some metaphysical entity who certainly did know his physics?
When thinking about \emph{the design argument}, it mostly goes something like this \cite{Colyvan2005}
\begin{enumerate}
    \item Our Universe seems remarkably fine-tuned for the emergence of carbon-based life;
    \item The boundary conditions and laws of physics could not have differed too much from the way they are in order for the Universe to contain life;
    \item Our Universe does contain life.
\end{enumerate}
Therefore,
\begin{enumerate}
    \item Our Universe is improbable;
    \item ``Our Universe is created by some intelligent being'' is the best explanation.
\end{enumerate}
Such that 
\begin{enumerate}
    \item A universe-creating intelligent entity exists.
\end{enumerate}
The jump to ``there is a god'' is not that far away anymore. Of course, this seems like the perfect evidence for Intelligent Design Creationism. 
By far, theists would likely adhere to this resolution as it gives and `intuitive' explanation and it `proves' the existence of an intelligent creator. Even when there is no evidence 
that this intelligent creator is the same being as their god. It also gives a teleological explanation to the crux: the constants are specifically chosen such that we as humankind may thrive.\\
\indent Nevertheless, throughout history we have seen that more and more phenomena which were first being classified as `divine intervention' made way for a natural, physical explanation. 
It is likely that the fine-tuning problem can also be solved without 
invoking a metaphysical reality. But, the difficulty of this problem should not be underestimated. If there is indeed a universal mathematical structure as stated by M.~Tegmark \cite{Tegmark2008}, then the fine-tuning problem should be explained by this theory.


\subsection{Going deeper into the UV}

An obvious solution to why some parameter has such a `weird' value would be a more fundamental theory which exactly describes why \cite{Wall2016}. However, this does not mean this new theory 
is exempted from its own fine-tuning problem. From this point of view, we are actually simply delaying the question/problem itself. If we would continue this path and continuously find increasingly more fundamental theories, we keep on delaying the question until 
we reach a dead end or finally arrive, at for instance, Tegmark's universal mathematical structure. Reaching a dead end would be undermining as this would be our final delay of the problem and we will need to look for another solution. Or, we can hope that if we had made another choice
during our way up to this dead end we would end up at another more fundamental theory. Eventually, like stated in the previous subsection, one would end up at Tegmark's universal theory and hope that it can finally resolve our issue.\\
\indent We should not leave this question open until we discover this final theory. 
By thinking about this problem and coming up with possible resolutions we may end up with a possible explanation and/or a more fundamental theory for one constant specifically, just like how naturalness guided the discovery of the charm quark.
Or, we would at least learn what a more fundamental theory should incorporate. Science is searching for new ideas and testing if they lead to sensible conclusions and predictions. There is always a good chance something unexpected may come our way when 
looking more thoroughly at a certain idea.


\subsection{The inherent failure of probability}
\epigraph{\emph{Once you eliminate the impossible, whatever remains, no matter how improbable, must be the truth.}}{-S.~Holmes}

\noindent Let us assume we have some constant $\alpha$ which must lie in some narrow range: $\alpha \in [\beta - \delta, \beta + \epsilon]$ in order for life to evolve \cite{Colyvan2005}.
Currently, we have no explanation why $\alpha = \beta$ exactly, so each value in that range is equally likely. If there are no constraints on the possible values $\alpha$ can take, it can take any value on the real line $\mathds{R}$. 
Using the Lebesgue measure, the probability of $\alpha \in [\beta - \delta, \beta + \epsilon]$ when $\alpha \in \mathds{R}$ is vanishingly small. Hence, very improbable, because it is not exactly $0$. The problem is that no matter how big 
we make $\delta$ and $\epsilon$, the probability will still remain $0$ as the whole real line gives an infinite value and that interval a finite one. Therefore, a life-permitting universe seems very improbable. This becomes more problematic if we take the product 
of such probabilities for all the constants we need. Clearly, this comes down to what a probability of $0$ actually means: improbable or impossible. This is the failure of \emph{logical possibility}: for every standard probability computation, the answer will be $0$.
How can we show that if the interval is small, the probability is low but not $0$?\\
\indent A solution may be invoking \emph{physical possibility}: maybe $\alpha$ cannot take any value in $\mathds{R}$ to begin with. There could be some constraints limiting the range of values $\alpha$ can take consistent with the laws of physics. But applying this extensively, one would end up 
with a probability of $1$, because in this limit the constraint would be set by experimental data. This is not what we want, we need to find a set large compared to our interval but small enough to give a non-zero result. But what is this set exactly? In summary: this essentially shifts the question from probability 
to this specific choice of set.\\
\indent Instead of a set, we can look at a distribution coming from some physical constraint. Such a distribution would need a little density over the life-permitting range to give a small result.
Again, this boils down to what distribution we should take.\footnote{Someone following the principle of naturalness would choose, e.g., a Gaussian distribution around some value of order 1.}\\
\indent Maybe this type of probability analysis is just wrong to begin with. The problem may be probability theory itself: any set with a finite measure gives a probability of $0$ against a set of infinite measure. Invoking probability theory to 
make this argument more rigorous and subsequently pointing out that probability theory is of little use is subversive. Sometimes this is called $\emph{the Measure Problem}$ \cite{hossenfelder2019screams}.\\
\\
\indent Maybe we should look at Bayesian statistics for a change. As an example, we will look at the hypothesis $H$ as being the design argument and the evidence $E$ as being fine-tuning. Then creationists would like to show that 
\begin{equation}
    P(H|E) > P(\neg H|E).
\end{equation}
Using the inverse probability law, we could rewrite this as 
\begin{equation}
    P(E|H) \frac{P(H)}{P(E)} > P(E | \neg H) \frac{P(\neg H)}{P(E)},
\end{equation}
so eventually, it suffices to show the following
\begin{equation}
    P(E|H) \cdot P(H) > P(E|\neg H) \cdot P(\neg H).
\end{equation}
However, for this to work we actually have to assume something: $P(E) \neq 0$. And we just saw that using the rules of probability we would get $0$.\\
\indent We could of course use Bayesian statistics in the case of technical naturalness and quantify the sensitivity of the IR parameters' dependence on the UV parameters \cite{hossenfelder2019screams}. 
Different models can then be compared with each other depending on the assumptions. For instance, do we assume a symmetry which protects the value when going to the IR from the UV or not.
But, again, we would obtain a shift of our question: instead of a distribution we would need to explain a certain set of priors/axioms.\\
\indent It seems like every try to analyse the problem using probability theory either fails or shifts the question. Even then, the probability for our Universe to have constants with those values remains very low. 
Naturally, this solution seems counterintuitive; that we are simply the `lucky ones', mainly because probability itself can be counterintuitive.


\subsection{Living in a Hubble Bubble or black hole}
A somewhat controversial topic is that of the \emph{Multiverse}, the claim that we live in a universe together with other possible universes causally disconnected from each other. Some theories actually predict this as a possible scenario.
For instance, inflation theory was proposed by A.~Guth \cite{Rijcke} to explain why our Universe is so flat. A consequence of this theory is \emph{eternal inflation}: the expansion rate of the Universe is modelled by the vacuum expectation value of a scalar field, the inflaton, in a certain potential. 
When this inflaton reaches a local minimum, inflation stops and we end up with a region looking locally like a FLRW-like universe. However, due to quantum fluctuations, there is a probability that the inflaton fluctuates upwards in the potential. So if inflation ends in a certain region, it may continue in another one. Hence, the result is that 
there is some fluid of Hubble volumes, spheres with the size of the observable Universe, where in some of those volumes inflation already ended and in other volumes inflation keeps on happening. Therefore, we end up with a mechanism creating a large amount of universes \cite{Rijcke, smolin2007scientific}.\\
\indent How many different universes can be created according to this mechanism, or how many microstates can produce the macrostate of the observable Universe? To answer this question, we compute the entropy $S$.
An upper and lower bound can be found making use of two assumptions \cite{Rijcke}:
\begin{enumerate}
    \item The upper limit is obtained by collapsing it to a black hole, this is known as \emph{the Bekenstein bound} \cite{PhysRevD.23.287}: $S \leq 10^{123}$;
    \item The lower limit is obtained by a proton estimate and assumptions about minimal structure formation: $S \geq 10^{77}$.
\end{enumerate}
This means there are between $e^{10^{77}}$ and $e^{10^{121}}$ possible universes all causally disconnected from each other. This is not a problem if the space in which this eternal inflation happens is infinitely large, which is the case in this theory.\\
\indent In which properties may these universes differ? A possibility is that they actually differ in their values for the fundamental parameters turning our fluid into a vast ensemble of universes. That we happen to live in one of many where it permits live would then be reminiscent of the anthropic principle. Unfortunately, this mechanism is not falsifiable due to the causal disconnectedness 
and also suffers from the measure problem.\\
\\
\indent L.~Smolin proposed that there might be a Popper falsifiable theory explaining the fine-tuning of the fundamental constants \cite{Smolin1994}. According to B.~DeWitt, when a star collapses to a black hole there is a probability that due to 
quantum fluctuations near the singularity where the densities become so extreme, the individual worldlines of the star's atoms start to diverge instead of converge. This region of diverging worldlines would be locally indistinguishable from an expanding universe with an apparent singularity in the past of 
every geodesic -- a Big Bang \cite{Smolin1994}. So, each new black hole creates a new universe.
Combine this with a proposal from J.~A.~Wheeler \cite{Misner2017}: the fundamental constants change at these initiations of new universes.\footnote{Such ideas are consistent with the String Theory Landscape, see Section \ref{section7}.}
L.~Smolin then proposed the following \cite{Smolin1994}:
\begin{center}
    All the dimensionless parameters
    change by small random values at such events, when black holes create new universes.
\end{center}
If these changes are thought of as genetic mutations \cite{ODowd2019} one ends up at \emph{Cosmological Natural Selection}: a universe prefers values for the fundamental constants such that the formation rate of black holes is maximised, because in that way their `cosmic genetics' are far more likely 
to be propagated. This might be an explanation for the values in our Universe. In his original paper \cite{Smolin1994} Smolin showed that if we were to change certain values like the weak or electromagnetic coupling constant, the black hole formation rate would decrease. 
\begin{center}
    Almost every small change in the parameters will either result in a universe that has less black holes than our present universe,
    or leaves that number unchanged.
\end{center}
Since black holes in our Universe are formed by collapsing massive stars, the black hole formation rate is sensitive to the rate at which those massive stars form, thus also on supernovae production and the existence of spiral galaxies. If our Universe allows a high star formation rate, there will be a lot of 
main sequence solar-like stars with planetary systems possibly admitting live. This idea can be thought of as giving a probability distribution with a peak around the values permitting a large black hole formation rate.\\
\indent What about changes that actually increase the black hole formation rate, can those happen? An example may be the baryon density in our Universe. If there are more baryons, there is more material available for star formation. However, this quantity can be related to the observed baryon-antibaryon asymmetry possibly stemming 
from CP violation which on itself it not entirely understood yet. Smolin proposed to look at the mass of the strange quark \cite{ODowd2019, Smolin1994}. \\
\indent Now consider the Chandrasekhar mass which is the cut-off mass between a star ending its life as a neutron star or as a black hole: a mass lower than this limit results in a neutron star, a higher mass would 
result in a black hole. This cut-off is sensitive to the equation of state describing nuclear matter which is itself described by QCD, the theory of the strong interaction. We can lower the cut-off mass by softening the equation of state which can be done by altering the mass of the strange quark. If we want optimal black hole formation, then the 
mass should be optimised to make the Chandrasekhar mass as low as possible. Smolin came to the conclusion that this would mean the cut-off mass lies at two solar masses. This is a Popper falsifiable prediction! From astronomical data we can deduce whether neutron stars with masses above two solar masses exist or not.
Unfortunately for cosmological natural selection, there has been an observation of a neutron star with a mass of about 2.17 solar masses last year \cite{Cromartie2020}. Even this year, there is a report about a possible neutron star of between 2.5 and 3 solar masses. This could be the lightest possible black hole ever found to date. 
There is no conclusion whether it is a neutron star or black hole yet \cite{Abbott2020}.\\
\indent A.~Vilenkin opposed this proposal by arguing that increasing the cosmological constant leads to a higher black hole formation rate \cite{Vilenkin2006}. His mechanism of black hole formation was through vacuum quantum fluctuations such that bigger universes create more black holes, although this requires 
a lot of dark energy which is not the case in our Universe. This suggestion also suffers from the measure problem. 
However, Smolin refutes these arguments since this requires assumptions of physics on extremely long time scales \cite{Smolin2006}.\footnote{More information about this principle, criticism and a possible theory favouring cosmological natural selection can be found in \cite{smolin2007scientific, Smolin2006, Dowker2017}.}\\
\\
\indent Even when this theory seems daunting and far-fetched, since it extrapolates an emergent biological principle which is rather complex on itself to cosmological scales, it is rather powerful since it actually allows falsifiable predictions. 
Nevertheless, these new ideas are the ones which try to probe the edges of our current understanding and we may learn new things about our Universe even if this theory is not the correct solution to the fine-tuning problem. This is what contributes to the progression of science.
However, this proposal also shifts the question towards how this mechanism would actually work. We still do not know everything about black holes and what their interiors look like, so it is still pretty speculative.\\
\indent Smolin assumed that the only production of black holes is due to stellar collapse. But there is also the possibility of primordial black holes \cite{Zeldovich1966}: black holes which already formed in the early Universe and can be seen as a dark matter candidate. Would such black holes also harbour 
universes and would a universe tend to have fundamental constants such that the formation of primordial black holes is optimised instead of stellar black hole formation?\\
\indent The proposal also refers to the coincidence between the elements making up life and the elements needed for star formation. Hence this coincidence seems to be based on the Goldilocks principle and our understanding of what life entails. 
But like previously stated, we do not know if there are other forms of life possible.\\
\\
\indent Black hole physics, especially the interior, still remains heavily theoretical as we do not have any means yet of doing experiments related to those objects. A theory describing the interior of black holes would give us a definite answer, but 
this is far from our current understanding. Recent developments include how we may reconstruct the interior from its Hawking radiation \cite{Almheiri2019, Penington2020}, so if we were able to capture and study Hawking radiation we could probe the interior and look directly at the answer.
This is due to the fact that this radiation is entangled with the collapsed matter which formed the black hole or any other matter which fell into it during its lifetime. Even so, we are not extracting the objects themselves but only the information so that we may recreate the objects. The objects themselves 
have a worldline which will forever be trapped inside the event horizon.\\
\indent Moreover, it is claimed that the universes inside the black holes are causally disconnected from the outside. But, we can still throw matter into the black hole altering its state. According to the \emph{No Hair theorem} black holes are uniquely defined by three parameters: its mass, electric charge and angular momentum $(M, Q, L)$ \cite{Carter1971, Israel1967, Israel1968}.
We could imagine throwing in an object with $(\delta M, \delta Q, \delta L)$, the black hole would then quickly settle to a new state with $(M + \delta M, Q + \delta Q, L + \delta L)$. How would this have an effect on the universe inside this `new' black hole? Would it be possible for an outsider to alter our Universe through such means?\\
\indent S.~Hawking's calculations showed that black holes emit Hawking radiation and hence evaporate and shrink \cite{Hawking1975}. Would such an evaporation be observable by someone in the interior and what would happen to the universe inside? Would this be described by a contracting universe?
Nonetheless, we do not know how the black hole will end: it could evaporate to zero leaving nothing but radiation, a Big Crunch-like scenario, or it could result in a remnant of a radius smaller than the Planck scale $10^{35}~\text{m}$ \cite{Strominger1995}.\footnote{The Planck scale is the scale at which general relativity and quantum field theory break down and quantum gravity is required.} 
Note that the evaporation rate highly depends on the black hole's initial mass and is much larger than the current age of the Universe for stellar black holes. This explains why our Universe, if trapped inside a black hole, is still intact and no effects have been observed yet.

\section{The Art of Science}
\label{section7}
Throughout its history, many physicists have used a concept to justify the choice of a specific model over another and dubbed it \emph{the Principle of Naturalness} \cite{borrelli2019practice}, although it is hardly based on any scientific principle.
Such models contain free parameters which take on more `beautiful' or `natural' values, generally values which are closer to the order of 1. 
They often argue that absurdly large or small values are being cherry-picked purely to correctly describe observations. 
However, one can easily reverse the argument and argue that values close to 1 are actually more cherry-picked than a random value.
In this sense, when faced with a new model one would deliberately assign some sort of distribution of values for the free parameters\footnote{This could be a Gaussian with variation 1.} which prefers more natural numbers over unnatural ones.\\
\indent In Tegmark's ``The Mathematical Universe'' \cite{Tegmark2008} he puts forward the idea of the \emph{Mathematical Universe Hypothesis}: there exists an independent external physical reality which is isomorphic to a mathematical structure. Furthermore, he argues that 
this is exactly what physicists mean by \emph{the Theory of Everything}. Such a theory would obviously contain fundamental free parameters from which all others can be derived. There is hence a good probability that this pure mathematical structure would contain unnatural numbers.
Moreover, if this theory is unique then we will never find a solution to the fine-tuning problem as this theory would need to incorporate possible solutions like natural cosmological selection. Even if someone believes in the design argument or anthropic principle but simultaneously also 
in Tegmark's Mathematical Universe Hypothesis, this omnipotent structure would need to describe these two views.\\
\indent The upholding of this view leads to the \emph{Uniqueness Problem} stated by E.~Wigner \cite{Wigner1960}: how can we be sure that this theory is unique. Throughout history we have found different models describing the same phenomena, e.g. the phlogiston theory was successful until it made way for 
a theory about combustion which could explain some gaps the phlogiston theory had. In this case, the discovery of O$_2$ led to a more fundamental theory,\footnote{A favourite example for pessimistic meta-induction.} but let us take a look at string theory. \\
\indent In \cite{Susskind2003} L.~Susskind argues that most physicists nowadays believe that the laws of Nature are uniquely described by some action principle determining the vacuum, the forces, the symmetries and the spectrum of elementary particles. One hence speaks of different solutions to some master theory 
(maybe Tegmark's hypothesis). The space of these solutions is called the \emph{moduli space}. But there is a problem, this cannot possible describe our Universe since all these solutions have massless particles and a vanishing cosmological constant. Susskind then argues that there should be something called 
the \emph{landscape}, the space of all possible vacua with a non-vanishing cosmological constant. He bases this claim on the observation that the cosmological constant is minute and absurdly fine-tuned such that it is almost impossible to find a vacuum in the observed range unless there is a humongous number of solutions with 
different values for the cosmological constant. But, according to Tegmark these should ultimately all be the same since there is only one universal mathematical structure. How can we determine which solution we have to choose? 
Suppose that these solutions are not the same and equally describe our Universe, then someone believing in Tegmark's hypothesis would conclude that this cannot possibly be the theory of everything.\\
\\
\indent Just like Wigner argued, by closing our eyes we could miss theories with the same strength and which may eventually lead to a more fundamental theory. Much like how one can take a different road at each intersection, the final destination can/will be immensely different had we taken another road somewhere along the way. Invoking the principle is hence 
the same as closing off some directions whilst we do not know whether those alternative routes could be better in the long run or not. This could in fact hinder the advancement of science. Of course, naturalness had some successes like the prediction of the charm quark, but even then one should not be blind to other possibilities. 
In today's state of physics, these alternative roads may be ideas like entropic gravity \cite{Verlinde2011}, emergent gravity \cite{Cao2017}, primordial black holes \cite{Zeldovich1966}, and so forth. Of course, according to Tegmark the final destination should be the same regardless of our choices. Even so, we can still end up in dead ends.\\
\indent Briefly put: this is a dangerous path to tread, as models should not be rejected purely because some constants have certain numerical values. Such a conclusion would hardly be based on any scientific methodology. Physical models and theories remain the best description of 
a natural phenomenon we currently have. Judging different models with the same predictive power purely on the numbers it contains would fade the line between what we call science and what we call art. As science is defined to be 
objective, we should refrain from any subjective arguments. Nonetheless, it seems that naturalness guided and still guides further developments in physics, especially in the branches of high energy physics and cosmology. 

\section{Conclusions}
\label{section8}
We discussed what we mean by constants when they appear in a model, and which ones are meaningful and which ones are not. In this context, the concepts of dimension and if they are derived or not played a key role. 
Furthermore, we have discussed how naturalness came to be, its uses and its successes but argued that a continued use of this `principle' may prove dangerous. Next, we described the fine-tuning problem and looked at the Goldilocks principle and its 
most important missing point: other forms of life. Subsequently, we looked at some physics examples and some proposed resolutions to the problem.\\
\indent Even when cosmological natural selection is the only candidate with falsifiable predictions, it still remains a highly non-trivial matter. Even this proposal suffers from something we also saw in other suggestions: a shift of the question we are trying to answer. In almost all proposals there was some 
apparent shift pushing the answer to questions about something else, some sort of \emph{truth shift}: what probability distribution we should take, what describes the black hole interior, etc. \\
\indent The debate about the fine-tuning problem is still very lively today and there are many more interesting articles to read, including those with more mathematical arguments or physical computations. Moreover, we could also discuss why we do not question the postulates a mathematical consistent theory is based on instead of focusing on the numbers \cite{hossenfelder2019screams}.

\section*{Acknowledgements}
This paper was written as part of the course ``History and Philosophy of Science: Physics \& Astronomy'' at Ghent University. I would like to thank the lecturers: Johan Braeckman, Maarten Boudry, Sven De Rijcke and Maarten Van Dyck for presenting interesting topics. 
It proves that taking up such a course is fruitful for any scientist as you learn to think more critically about the meaning of science instead of only doing computations. Knowing its history is a valuable asset for future progression. Special thanks goes to Johan Braeckman for giving me his thoughts on earlier drafts, 
his insightful comments helped shape the paper, and his advice to try to publish this as a paper.


\bibliography{Finetuning_references.bib}

\end{document}